\documentclass[pdflatex,sn-mathphys-num]{sn-jnl}

\usepackage{graphicx}%
\usepackage{multirow}%
\usepackage{amsmath,amssymb,amsfonts}%
\usepackage{amsthm}%
\usepackage{mathrsfs}%
\usepackage[title]{appendix}%
\usepackage{xcolor}%
\usepackage{textcomp}%
\usepackage{manyfoot}%
\usepackage{booktabs}%
\usepackage{algorithm}%
\usepackage{algorithmicx}%
\usepackage{algpseudocode}%
\usepackage{listings}%
\usepackage[T1]{fontenc} 
\usepackage[utf8]{inputenc}
\usepackage{geometry}
\usepackage{booktabs}   
\usepackage{tabularx}   
\usepackage{ragged2e}   
\usepackage{url}        

\usepackage{lineno}
\usepackage[english]{babel}
\usepackage{csquotes}

\theoremstyle{thmstyleone}%
\theoremstyle{thmstyletwo}%

\theoremstyle{thmstylethree}%

\begin{document}

\title[Article Title]{The Rise of AI in Weather and Climate Information and its Impact on Global Inequality}


\author*[1]{\fnm{Amirpasha} \sur{Mozaffari}}\email{amirpasha.mozaffari@bsc.es}
\author[1]{\fnm{Amanda} \sur{Duarte}}
\author[1]{\fnm{Lina} \sur{Teckentrup}}
\author[1]{\fnm{Stefano} \sur{Materia}}
\author[2]{\fnm{Gina E. C.} \sur{Charnley}}
\author[1]{\fnm{Lluís} \sur{Palma}}
\author[1]{\fnm{Eulalia} \sur{Baulenas Serra }}
\author[1]{\fnm{Dragana} \sur{Bojovic}}
\author[1]{\fnm{Paula} \sur{Checchia}}
\author[3]{\fnm{Aude} \sur{Carreric}}
\author[1,4]{\fnm{Francisco} \sur{Doblas-Reyes}}



\affil*[1]{\orgdiv{Earth Sciences Department}, \orgname{Barcelona Supercomputing Center (BSC)}, \orgaddress{\street{Plaça d'Eusebi Güell, 1-3}, \city{Barcelona}, \postcode{08034}, \country{Spain}}}

\affil[2]{\orgdiv{School of Public Health}, \orgname{Imperial College London}, \orgaddress{\street{White City Campus}, \city{London}, \postcode{W12 7TA}, \country{United Kingdom}}}

 \affil[3]{\orgdiv{Independent Researcher}}

\affil[4]{\orgname{Catalan Institution for Research and Advanced Studies (ICREA)}, \orgaddress{\street{Passeig Lluís Companys 23}, \city{Barcelona}, \postcode{08010}, \country{Spain}}}


\abstract{The rapid adoption of AI in Earth system science promises unprecedented speed and fidelity in the generation of climate information. However, this technological prowess rests on a fragile and unequal foundation: the current trajectory of AI development risks further automating and amplifying the North-South divide in the global climate information system. We outline the global asymmetry in  High-Performance Computing and data infrastructure, demonstrating that the development of foundation models is almost exclusively concentrated in the Global North. Using three different domains, we show how this infrastructure inequality continues through models’ inputs, processes and outputs. As an example, in weather and climate modelling, the reliance on historically biased data leads to systematic performance gaps that disproportionately affect the most vulnerable regions. In climate impact modelling, data sparsity and unrepresentative validation risk driving misleading interventions and maladaptation. Finally, in large language models, dependence on dominant textualised forms of climate knowledge risks reinforcing existing biases. We conclude that addressing these disparities demands revisiting the three phases, i.e. models’ \enquote{Input}, \enquote{Process} and \enquote{Output}. This involves  (i) a perspective shift from model-centric to data-centric development, (ii) the establishment of a Climate Digital Public Infrastructure and human-centric evaluation metrics, and (iii) a move from producer-consumer dynamics toward knowledge co-production. This integration of diverse knowledge systems would truly democratise compute sovereignty and ensure that the AI revolution fosters genuine systemic resilience rather than exacerbating inequity.}





\maketitle

\section{Introduction}\label{sec1}

The \enquote{quiet revolution} of numerical weather prediction \citep{bauer2015quiet} is rapidly giving way to a transformative era of Artificial Intelligence (AI) and Deep Learning in Earth system science \citep{reichstein2019deep}. These technologies promise to resolve long-standing bottlenecks in process-based modelling: they offer hybrid solutions for sub-grid-scale parameterisation \citep{eyring2024ai, couvreux2021process}, extend observational records through reconstruction \citep{plesiat2024artificial, guo2025reconstructed}, and dramatically reduce the computational cost of forecasting, thereby extending prediction horizons \citep{bracco2025machine}. Beyond standard forecasting, these tools are being explored for climate change and variation attribution mapping  \citep{callaghan2021machine}, and for their potential in detecting tipping points in the Earth System \citep{huang2024deep}. 

At the same time, large language models (LLMs) are reshaping how climate knowledge is synthesised, accessed, and operationalised. By mediating access to scientific assessments, technical documentation, and policy-relevant information, LLMs extend the influence of AI in Earth system science.
However, this technological prowess rests on a fragile foundation. Existing AI systems frequently rely on geographically skewed training datasets and concentrated infrastructure. 

As McGovern et al. \citep{IdentifyingandCategorizingBiasinAIMLforEarthSciences} warn, this reliance risks entrenching 'systemic and structural bias,' where historical disparities in observation networks, such as the relative scarcity of ground-based sensors in the Global South, are not only preserved but automated, effectively perpetuating environmental injustice. Biased climate information has the potential to destabilise political institutions by informing inequitable adaptation policies, thereby jeopardising social equilibrium and potentially precipitating societal tipping points \citep{debnath2023harnessing}. Advancements in AI-powered weather prediction have raised the claim that these methods are democratising weather predictions (e.g., \citep{hall2025ai}, \citep{azhar2025fully}, \citep{khadir2025democracyainumericalweather}), as a similar claim appears regarding climate services knowledge with the increased popularity of LLMs (e.g., \citep{kuznetsov2025transforming} ).
As the AI for a Planet Under Pressure study \citep{galaz2025ai} highlights, science plays a critical role in ensuring that AI development does not exacerbate planetary pressures or deepen existing inequalities. This paper examines the intersection of AI, climate science, and global inequality, arguing that without a deliberate restructuring of how we curate data and deploy infrastructure, the AI revolution risks leaving the most vulnerable behind.

\section{Manifestations of Inequality}\label{sec2}

AI systems can be understood as socio-technical systems that transform inputs into outputs through computational processes. Inputs include the goals defined for the system and the data on which it relies, encompassing both the selection of tasks to be solved and the collection of observational, simulated, or human-generated data. The process stage comprises data pre-processing, model (pre)training, and evaluation, during which data are cleaned and structured, model parameters are learned or fine-tuned using computational infrastructure, and performance is assessed against chosen metrics and benchmarks, which affects back the inputs. Outputs consist of the model’s predictions, classifications, or recommendations, as well as the ways these outputs are deployed, accessed, and integrated into decision-making contexts.

\begin{figure}
  \centering
  \includegraphics[width=0.99\linewidth]{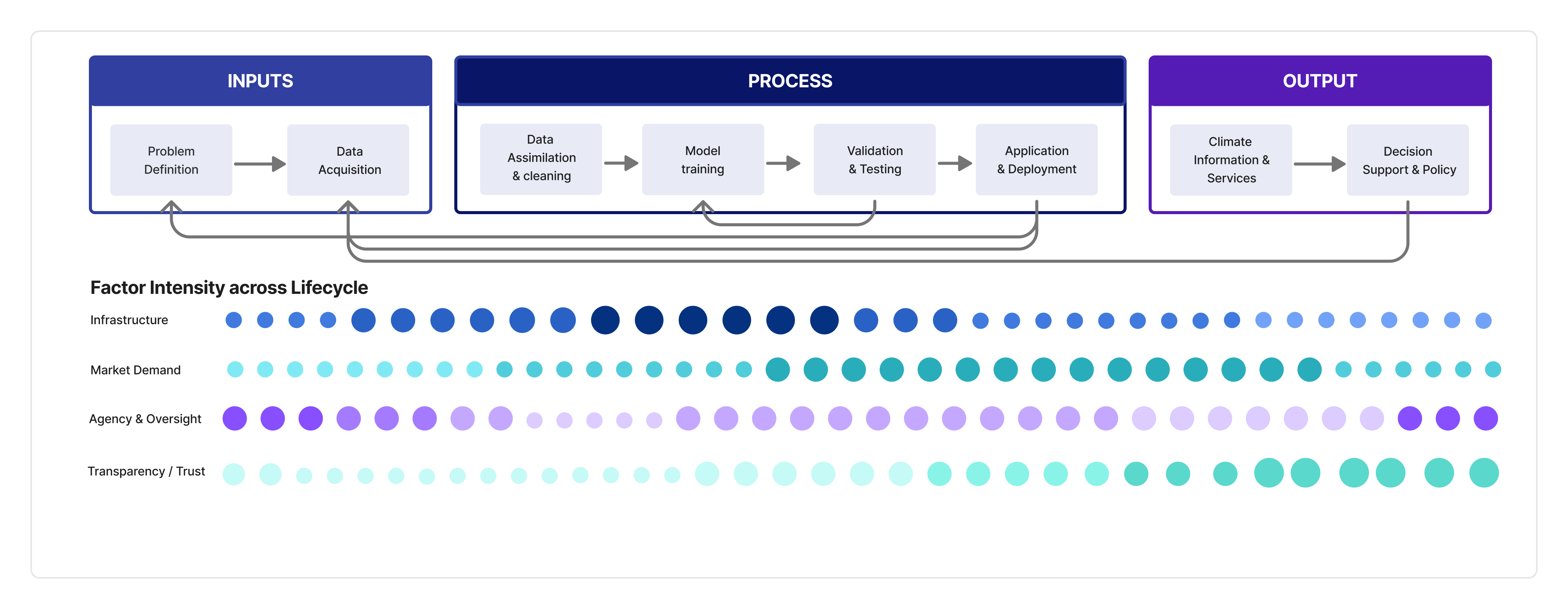}
  \caption{The lifecycle of AI model development in weather and climate sciences, tracing the workflow from initial data acquisition to downstream application. The bottom panels qualitatively illustrate the varying intensity of four key dimensions across these stages: the level of human agency and oversight, the degree of transparency and trust, the magnitude of infrastructure requirements, and the existence of market demand.} 
  \label{fig:flowchart} 
\end{figure}

These steps are iterative and have fuzzy boundaries, but across them, biases are accumulated, producing a cascading effect towards exacerbating global inequalities. In the following sections, we examine how such biases emerge across the three main stages in AI applications for weather, climate, and related domains.

\subsection{Input Level}\
The first and main entry point for biases is at the input stage. Research agendas are shaped largely by institutions and funders based in the Global North, which tend to prioritise a technocratic framing of climate challenges, emphasising efficiency, prediction accuracy, and high-technology solutions over equity-oriented or locally embedded forms of climate knowledge \citep{overland2022funding}. This already impacts the goal and problem definition, upon which AI models will be based.

In terms of data acquisition, reanalysis datasets, which blend observations with past short-range weather forecasts, rerun with modern forecasting models serve as the cornerstone of AI-based weather and climate prediction due to their spatial and temporal consistency. As Table \ref{tab:frontier_models_appendix} shows, most frontier AI models are pre-trained or fine-tuned on ERA5 \citep{hersbach2020era5}, widely considered the gold standard for historical atmospheric data. However, this reliance on \enquote{maps without gaps} creates an illusion of uniform global quality. While variables such as temperature and pressure are directly assimilated from observations, the density of these inputs is highly unequal; in observation-sparse regions, the reanalysis is significantly less constrained by real-world data. Furthermore, other variables (most notably precipitation) are generated by internal model physics rather than direct measurements. This creates a \enquote{model-of-a-model} dependency; trained systems inadvertently inherit and reinforce the latent biases of the numerical models used to produce their training data. However, ERA5 has well-documented biases across multiple regions and variables.  Notable, yet far from exhaustive, examples include discrepancies in precipitation compared to observational datasets \citep{jiao2021evaluation} and systematic lower agreement with observations in tropical regions, including South America, Central Africa, and Southeast Asia \citep{lavers2022evaluation}. In the Amazon, these discrepancies stem from two issues: first, ERA5’s reliance on static land-use maps fails to capture deforestation impacts, underestimating climatic trends compared to ground stations \citep{da2024discrepancies}. Second, reanalysis products report significant drying patterns in the southern and central Amazon that are absent in observational records \citep{polasky2025discrepancies} \citep{cattelan2025unraveling}. Similar biases manifest as air temperature errors in China \citep{zou2022performance}, and inaccuracies in interannual and decadal precipitation variability across the Mongolian Plateau \citep{xin2022evaluation}. Likewise, systematic errors in air temperature have been detected in the Ethiopian highlands \citep{christensen2026errorera52mtemperature}. This reliance on simulated data is mirrored in AI development for climate, which depends on climate model projections such as CMIP6 \citep{eyring2016overview}. Ultimately, this dependence on assimilated (ERA5) and simulated (CMIP6) datasets will likely persist until AI models trained purely on observational data ( e.g. \citep{allen2025endtoend}) can demonstrate comparable predictive performance.

This observational inequality extends beyond atmospheric physics into the biosphere and ecosystem services. For instance, the FLUXNET2015 \citep{pastorello2020fluxnet2015} is an extremely valuable dataset that provides ecosystem-scale data on CO$_2$, water, and energy exchange between the biosphere and the atmosphere, yet the spatial distribution of flux towers is highly concentrated in Europe and North America. Similarly, Europe’s well-resourced Integrated Carbon Observation System (ICOS) contrasts starkly with the resource-constrained West African Flux Network (WAFNET), while the Global Greenhouse Gas Reference Network (GGGRN) primarily focuses on the mainland United States. Even the community-driven FLUXNET-CH4 \citep{delwiche2021fluxnet} datasets, released in 2021, suffer from the same disproportionate representation of the advanced Global North. Consequently, widespread usage of these datasets in ML methods (e.g., FLUXCOM\citep{jung2019fluxcom}, MetaFlux\citep{nathaniel2023metaflux}, GloFlux\citep{yuan2025global}) risks embedding these geographic blind spots into the next generation of carbon cycle models.

In many AI applications, climate and environmental data do not operate in isolation but are combined with datasets from related sectors, where additional biases may arise. Climate–health studies provide a clear example, in which weather and climate information is integrated with health data that suffer from underreporting, over-reporting and pervasive biases: healthcare access varies, testing is inconsistent, case definitions change over time, and surveillance tends to focus on easy-to-reach populations or diseases which are a current health priority \citep{wong2023leveraging}. Together, these factors make it extremely difficult to estimate the true global burden of many diseases, particularly infectious diseases which largely impact populations in extreme poverty and introduce substantial uncertainty into AI models trained on such data, leading to out-of-sample failures \citep{kraemer2025artificial}.

Finally, in recent years, several climate and Earth science-specific LLMs have emerged. These models rely on large text corpora scraped from the internet \citep{marion2023when}\citep{zewang2025bias} and are typically fine-tuned on scientific literature, environmental reports, or specialised datasets \citep{lin2024designing}. These include authoritative sources such as the Intergovernmental Panel on Climate Change (IPCC) assessment reports, the Intergovernmental Science-Policy Platform on Biodiversity and Ecosystem Services (IPBES) or the Lancet Countdown on Health and Climate Change. These reports synthesise research outputs produced, among others, by the upstream applications discussed above. Examples include ChatClimate, which grounds GPT-4 responses in the IPCC Sixth Assessment Report; ClimateQ\&A, which integrates a knowledge base derived from IPCC and IPBES documents; ClimateGPT, trained on interdisciplinary climate literature; or ChatNetZero, a retrieval-augmented system focused on climate pledges \citep{kuznetsov2025transforming}. While such models aim to improve the accuracy and relevance of climate reasoning, they may reproduce and even amplify the longstanding geographic and linguistic imbalances present in their source materials \citep{pasgaard2015geographical}\citep{IdentifyingandCategorizingBiasinAIMLforEarthSciences}\citep{manvi2024large}\citep{sanu2024limitations}\citep{zewang2025bias}.

As shown in the bottom panel of Fig. \ref{fig:flowchart}, the 'Input' phase features low market demand and minimal infrastructure requirements, combined with high agency in agenda-setting and transparency.

\subsection{Process Level}\

Applying AI to climate and other climate-impacted domains comes with structural challenges that arise before model training. Different domains generate, curate, and store data in fundamentally different ways, meaning that harmonising them often requires substantial aggregation or other forms of coarse pre-processing. These early integration choices shape the information that ultimately becomes available for model training and evaluation.

For model training and evaluation, High-Performance Computing (HPC) infrastructure, the engine of modern weather and climate science \citep{bauer2015quiet}, is key. These facilities not only run operational forecast systems but also serve as the primary repositories for the petabytes of training data required by modern AI. However, access to this \enquote{compute} is profoundly unequal. As detailed in Appendix Table \ref{tab:hpc_countries_appendix}, the major publicly funded HPC systems—those that form the backbone of global operational forecasting—are heavily concentrated in the Global North and East Asia \citep{zaidan2025artificial}. Consequently, frontier AI-weather models are almost exclusively developed through collaborations between these national public centres and a handful of multinational technology firms (e.g., Google DeepMind, NVIDIA, Microsoft, Huawei). This concentration effectively gatekeeps the development of foundation models, restricting it to entities with access to exascale computing and specialised engineering talent (see \enquote{Infrastructure} requirement in Fig. \ref{fig:flowchart}). 

While the computational cost of deploying these models is significantly lower than training them, the infrastructure gap remains a barrier to equitable access. Currently, only 32 countries globally host the hyperscale AI data centres necessary for robust cloud-based deployment \citep{satariano2025aicompute}. A stark resource divide binds the physical realisation of AI. Data centre energy demands are projected to reach 4.5\% of global generation by 2030, the pressure on local grids is becoming acute \citep{nature2025energy}. An equally voracious demand for water matches this thirst for power: by 2027, global AI demand is projected to account for 4.2--6.6 billion cubic meters of water withdrawal \citep{li2023making}. The substantial capital and resource demands of this infrastructure place it out of reach for much of the Global South. This creates a risk of technological dependency, where low-income nations are relegated to being consumers of models that are not generated for them as the main target by infrastructure they cannot audit, control, or adapt to their local contexts. Without compute sovereignty \citep{hawkins2025ai}, the ability to train and fine-tune models on sovereign infrastructure, nations lack the agency to prioritise their specific regional climate vulnerabilities \citep{regilme2024artificial}.

The reliance on biased training datasets, whether in weather, ecosystem, health or language-based AI applications, has direct implications for model benchmarking and evaluation. Benchmark datasets and challenges are essential for comparing machine learning tools and approaches \citep{dueben2022challenges}, and a growing number of benchmark datasets have been developed for climate and weather applications (Table \ref{tab:benchmark_datasets_appendix}).  Similarly, as it happens at the input level, these benchmarks often rely on the same underlying datasets, most notably ERA5 and CMIP6, thereby inheriting their limitations. In the case of LLMs, evaluation is similarly shaped by upstream choices: quality-filtering pipelines frequently reflect Western linguistic or cultural assumptions, privileging hegemonic viewpoints \citep{bender2021dangers}, while benchmark and ground-truth datasets remain unevenly representative across regions \citep{pasgaard2015geographical}. Together, these factors risk propagating and legitimising existing biases throughout the evaluation pipeline. Furthermore, the optimisation process itself embeds implicit biases. As training regimes typically employ gradient descent to minimise Root/ Mean Squared Error (R/MSE), the steepest descent in the global loss function is determined by data-dense regions, predominantly the Global North. Consequently, the algorithm prioritises error reduction in well-observed latitudes, treating data-sparse regions as negligible. Furthermore, minimising R/MSE drives models toward the conditional mean, generating \enquote{blurry} forecasts that smooth out critical extreme events to prioritise global average statistics over local physical fidelity.

Finally, sovereignty relies on the unhindered flow of information, a form of “soft infrastructure” that is increasingly threatened by commercialisation and funding cuts. As Taalas \citep{doi:10.1126/science.aec8146} warns, restricting access to publicly funded climate data would disproportionately harm low-income nations that lack the financial means to purchase essential climate information. These challenges are compounded by access constraints at the level of AI deployment. Limited or unequal internet connectivity poses an additional barrier to use \citep{bender2021dangers} \citep{unpress2023gaef3587}, while key capabilities of frontier LLMs are often locked behind pay-walled APIs or proprietary cloud platforms, placing them beyond the reach of many low-resource institutions.

\subsection{Output Level}\

Climate information and services play a central role in supporting decision-making across climate-sensitive sectors, informing policy design, risk management, and long-term adaptation planning. Advances in Earth-system science and data-driven methods have expanded the range of climate products available to governments, public agencies, and practitioners, from forecasts and early-warning systems to impact assessments and decision-support tools. These services are increasingly expected not only to improve predictive accuracy but also to translate scientific knowledge into actionable insights that can guide policy interventions and resource allocation. However, the potential of climate information systems to become supportive of decision-making depends critically on how underlying data, models, and AI-driven tools are developed, evaluated, and deployed, raising important questions about whose risks are prioritised and whose decisions these systems ultimately serve.

Such limitations intersect with a deeper equality problem in forecasting capabilities. Despite continuous improvements in forecast skill over recent decades, a meaningful gap persists across income levels. \citet{linsenmeier2023global} found that temperature forecasts are substantially more accurate in high-income countries, to the extent that a seven-day-ahead forecast in wealthy nations outperforms a one-day-ahead forecast in low-income countries. Concerningly, this trend continues with AI-powered models. A recent comparison of ECMWF's AIFS model with IFS HRES revealed that while AIFS offers significant overall gains in forecast skill, these benefits are unevenly distributed: wealthier and more densely populated regions see greater improvements, raising serious concerns about fairness and equity in AI weather prediction \citep{olivetti2025weather}.
Gaps in biogeochemical data have many impacts, from misunderstanding processes to misleading policymakers \citep{metcalfe2025gaps}, and lead to an over-representation of the extratropical Northern Hemisphere’s fingerprint in climate change \citep{devries2023robust} \citep{cattelan2025unraveling}.

Although AI-driven early warning systems and predictive dashboards are widely promoted, their real-world value is debatable—especially in the context of global inequality, where there is almost no rigorous evidence that they improve population outcomes, such as health \citep{MADDAH2023} \citep{Schulze2023}. For instance, many preventable infectious diseases disproportionately affect children under five living in poverty \citep{roser2024}. In such contexts, sophisticated warning systems do little to address fundamental drivers of vulnerability. Just as high-precision weather forecasts offer little protection to those lacking resilient shelter, disease predictions are of limited use without the means to intervene. Consequently, investments in basic public health infrastructure may yield far greater impact. 

In this field of AI-driven early warning systems and decision-making support, most of the AI applications rely on a narrow subset of techniques, particularly random forests and boosted regression trees. These methods are popular because they are accessible, robust, and tend to perform well on noisy data; however, there has been surprisingly little exploration of whether their dominance reflects genuine superiority for modelling disease transmission, or simply convenience and convention \citep{keshavamurthy2022predicting}. Overall, when AI is applied to support decision-making, it is often used as a black-box prediction tool. While such models can achieve high performance metrics, they rarely offer insight into the mechanisms or relationships underpinning climate risks. Without deeper interpretability, predictions may be misleading, fail out-of-sample, or provide limited guidance for understanding risk drivers, designing interventions, or supporting climate-adaptation planning \citep{brownstein2023advances}. One way to improve interpretability is through hybrid approaches that combine the flexibility of machine learning with mechanistic and statistical models \citep{kraemer2025artificial}, as well as through transdisciplinary collaborations, as discussed in the next section. In parallel, large language models (LLMs) are increasingly incorporated into climate information services as interfaces for synthesising, communicating, and operationalising climate knowledge. By mediating access to forecasts, impact assessments, and scientific evidence, LLMs shape how climate information is interpreted and used in decision-making and policy processes. As these models become more embedded in climate services, careful attention is needed to understand how they influence access, representation, and authority in climate-related decisions.

\section{A Call to Action: Forging an Equal and Equitable Path Forward}\label{sec3}

The claim that AI-powered predictions and related tools have a democratising effect for Earth system sciences is largely grounded in an instrumental conception of democracy. From this perspective, such tools are valued for their ability to increase the production of data and analytical capacity, often at a presumed lower cost. Although the access to and benefits of these tools are unevenly distributed, the overall gains in efficiency and decision-making capacity are assumed to outweigh these inequalities. However, if democracy is understood in intrinsic terms, namely, as valuable in itself for fostering equal conditions and participation, all the dynamics discussed in this paper call into question the purported democratising power of AI. We have seen how biases accumulate and create cascading effects across multiple stages of the modelling pipeline, compounding a prior injustice \citep{sambasivan2021everyone} \citep{hellman2023big}. If in an AI-driven world, knowledge authority is assigned to those who have a long-term record of developing technologies, models, or scientific publications and possess the infrastructure to build on them, we will perpetuate the deficient linear approach to producing new knowledge. In this scenario, technologically “powerful” ones are the “producer”, and the rest of the community is the “user” of new knowledge, ignoring the value of knowledge plurality and equal rights that real democracy should foster.

To navigate this ethical deficit, the community must move beyond performance metrics to adopt the FATES principles (Fairness, Accountability, Transparency, Ethics, and Sustainability) \citep{reichstein2025early} in every facet of the AI adoption in climate sciences, addressing the deficits in agency and transparency highlighted in Fig. \ref{fig:flowchart}. Operationalising these principles requires dismantling the infrastructure of inequality. Rather than proprietary silos, we need a unified, open-source ecosystem that functions as a digital public utility, i.e. we should move towards a Climate Digital Public Infrastructure. This 'soft infrastructure' ensures that vital climate information remains a global public good \citep{lemos2025digital}. However, achieving this requires overcoming a stark market misalignment illustrated in Fig. \ref{fig:flowchart}: demand surges for downstream applications but remains negligible for the foundational, yet capital-intensive, task of data acquisition. Encouraging mechanisms that incentivise investment in the upstream observational infrastructure would move the spotlight from model development and acknowledge the diminishing returns of prioritising compute infrastructure \citep{hoffmann2022training} above all else. While expanding observational networks is inherently slower and more arduous than scaling hardware, it offers a more sustainable pathway toward substantive improvements in model fidelity \citep{eyre2022value}.

This structural reform must be matched by a paradigm shift from model-centric to data-centric AI, redirecting focus from architectural novelty to the enhancement of data quality and representativeness \citep{zha2025data}. We envision an AI ecosystem for Earth science that shifts its optimisation targets from abstract statistical aggregates to the tangible safety and well-being of human populations. This requires a fundamental redesign of our evaluation metrics to prioritise the \enquote{human cost} of errors. In the context of weather and climate prediction, this means addressing the double penalty issue \citep{subich2025fixingdoublepenaltydatadriven}. Current objective functions often push models toward \enquote{blurrier}, smoothed predictions to minimise average global loss, obscuring the extreme local events that threaten lives. To counter this, the field must adopt fairness-aware loss functions \citep{olivetti2025weather} and population-weighted metrics \citep{hanigan2006comparison} that explicitly penalise forecast failures in densely populated or vulnerable regions, rather than treating far-ocean locations and urban centres as not only mathematically, but also risk-wise equivalent. Furthermore, fairer benchmark datasets and benchmark challenges will help promote and advance these goals \citep{masi2025safe}.

Ultimately, a fundamental shift in how we collaborate should accompany these technical advancements. In the climate services domain, this means adopting knowledge co-production frameworks that recognise a plurality of knowledge systems, Indigenous, and operational alongside technical expertise. By creating a “meeting place” \citep{bremer2017co} where diverse stakeholders collaborate \citep{turnhout2020politics}, we can dismantle the power hierarchy that privileges model developers over model users. Such partnerships are essential for generating AI-driven insights that are genuinely fit-for-purpose and democratic. The technological prowess of AI offers unprecedented opportunities to navigate the planetary crisis, but speed cannot come at the expense of equity. If we continue to build AI systems on skewed data and concentrated infrastructure, we risk automating the very environmental injustices we seek to resolve. However, by restructuring our digital foundations to be open and inclusive, and by centring the voices of the most vulnerable in our evaluation metrics, we can steer this revolution toward more just outcomes. The goal of AI in Earth science must not merely be the optimisation of loss functions, but the preservation of a livable planet for all.

\backmatter


\bmhead{Acknowledgements}

The illustration in this work was created by Diana Urquiza, whose contribution we gratefully acknowledge.

“AM acknowledges the Grant JDC2023-051208-I, funded by MICIU/AEI/10.13039/501100011033 and, as appropriate, by “ERDF A way of making Europe”, by “ERDF/EU”, by the “European Union”, or by the “European Union NextGenerationEU through PRTR”. “AD, SM, and TL, DB acknowledge their AI4S fellowship within the “Generación D” initiative by Red.es, Ministerio para la Transformación Digital y de la Función Pública, for talent attraction (C005/24-ED CV1), funded by NextGenerationEU through PRTR”. This work received funding from the European Union's Horizon Europe Framework Programme through the project CONCERTO (Grant Agreement 101185000). This work, as part of TerraDT - Digital Twin of Earth System for Cryosphere, Land Surface and related interactions, has received funding from the European Union’s Horizon Europe Framework Programme (HORIZON) under Grant Agreement no. 101187992. This work was supported by the European Union’s Horizon Europe research and innovation programme under grant agreement No 101214398 (ELLIOT). This work was supported by the European Union’s Horizon Europe research and innovation programme under grant agreement No 101056933 (Climateurope2)

\bmhead{Author Contributions}
A.M. conceptualised the work in early discussions with A.C., led the writing of the manuscript, and coordinated the research team. A.D., L.T., S.M., L.P., P.C., and G.E.C.C. contributed significantly to the drafting of the "Input" and "Process" sections. P.C., E.B.S., and D.B. focused on the structure of Section 2 and the "Output". F.D.R. provided overall supervision, strategic guidance, and critical revision of the manuscript. All authors reviewed and approved the final manuscript.

\bmhead{Competing Interests}
The authors declare no competing interests.






\bibliography{sn-bibliography}

@article{bauer2015quiet,
  title={The quiet revolution of numerical weather prediction},
  author={Bauer, Peter and Thorpe, Alan and Brunet, Gilbert},
  journal={Nature},
  volume={525},
  number={7567},
  pages={47--55},
  year={2015},
  publisher={Nature Publishing Group UK London}
}

@article{reichstein2019deep,
  title={Deep learning and process understanding for data-driven Earth system science},
  author={Reichstein, Markus and Camps-Valls, Gustau and Stevens, Bjorn and Jung, Martin and Denzler, Joachim and Carvalhais, Nuno and Prabhat, F},
  journal={Nature},
  volume={566},
  number={7743},
  pages={195--204},
  year={2019},
  publisher={Nature Publishing Group UK London}
}

@article{kochkov2024neural,
  title={Neural general circulation models for weather and climate},
  author={Kochkov, Dmitrii and Yuval, Janni and Langmore, Ian and Norgaard, Peter and Smith, Jamie and Mooers, Griffin and Kl{\"o}wer, Milan and Lottes, James and Rasp, Stephan and D{\"u}ben, Peter and others},
  journal={Nature},
  volume={632},
  number={8027},
  pages={1060--1066},
  year={2024},
  publisher={Nature Publishing Group UK London}
}

@article{bonev2025fourcastnet,
  title={FourCastNet 3: A geometric approach to probabilistic machine-learning weather forecasting at scale},
  author={Bonev, Boris and Kurth, Thorsten and Mahesh, Ankur and Bisson, Mauro and Kossaifi, Jean and Kashinath, Karthik and Anandkumar, Anima and Collins, William D and Pritchard, Michael S and Keller, Alexander},
  journal={arXiv preprint arXiv:2507.12144},
  year={2025}
}

@article{eyring2024ai,
  title={AI-empowered next-generation multiscale climate modelling for mitigation and adaptation},
  author={Eyring, Veronika and Gentine, Pierre and Camps-Valls, Gustau and Lawrence, David M and Reichstein, Markus},
  journal={Nature Geoscience},
  volume={17},
  number={10},
  pages={963--971},
  year={2024},
  publisher={Nature Publishing Group UK London}
}

@article{couvreux2021process,
  title={Process-based climate model development harnessing machine learning: I. A calibration tool for parameterization improvement},
  author={Couvreux, Fleur and Hourdin, Fr{\'e}d{\'e}ric and Williamson, Daniel and Roehrig, Romain and Volodina, Victoria and Villefranque, Najda and Rio, Catherine and Audouin, Olivier and Salter, James and Bazile, Eric and others},
  journal={Journal of Advances in Modeling Earth Systems},
  volume={13},
  number={3},
  pages={e2020MS002217},
  year={2021},
  publisher={Wiley Online Library}
}

@article{bracco2025machine,
  title={Machine learning for the physics of climate},
  author={Bracco, Annalisa and Brajard, Julien and Dijkstra, Henk A and Hassanzadeh, Pedram and Lessig, Christian and Monteleoni, Claire},
  journal={Nature Reviews Physics},
  volume={7},
  number={1},
  pages={6--20},
  year={2025},
  publisher={Nature Publishing Group UK London}
}

@article{huang2024deep,
  title={Deep learning for predicting rate-induced tipping},
  author={Huang, Yu and Bathiany, Sebastian and Ashwin, Peter and Boers, Niklas},
  journal={Nature Machine Intelligence},
  volume={6},
  number={12},
  pages={1556--1565},
  year={2024},
  publisher={Nature Publishing Group UK London}
}

@article{callaghan2021machine,
  title={Machine-learning-based evidence and attribution mapping of 100,000 climate impact studies},
  author={Callaghan, Max and Schleussner, Carl-Friedrich and Nath, Shruti and Lejeune, Quentin and Knutson, Thomas R and Reichstein, Markus and Hansen, Gerrit and Theokritoff, Emily and Andrijevic, Marina and Brecha, Robert J and others},
  journal={Nature climate change},
  volume={11},
  number={11},
  pages={966--972},
  year={2021},
  publisher={Nature Publishing Group UK London}
}

@misc{christensen2026errorera52mtemperature,
      title={Error in ERA5 2m Temperature identified using GraphCast}, 
      author={Hannah M. Christensen and Jack Barker and Bobby Antonio and Massimo Bonavita and Mohamed Dahoui and Patricia de Rosnay},
      year={2026},
      eprint={2601.04701},
      archivePrefix={arXiv},
      primaryClass={physics.ao-ph},
      url={https://arxiv.org/abs/2601.04701}, 
}

@misc{debnath2023harnessing,
  title={Harnessing human and machine intelligence for planetary-level climate action. npj Climate Action, 2 (1), 20},
  author={Debnath, Ramit and Creutzig, F and Sovacool, BK and Shuckburgh, E},
  year={2023},
  publisher={DOI}
}

@article{reichstein2025early,
  title={Early warning of complex climate risk with integrated artificial intelligence},
  author={Reichstein, Markus and Benson, Vitus and Blunk, Jan and Camps-Valls, Gustau and Creutzig, Felix and Fearnley, Carina J and Han, Boran and Kornhuber, Kai and Rahaman, Nasim and Sch{\"o}lkopf, Bernhard and others},
  journal={Nature Communications},
  volume={16},
  number={1},
  pages={2564},
  year={2025},
  publisher={Nature Publishing Group UK London}
}

@article{galaz2025ai,
  title={AI for a Planet Under Pressure},
  author={Galaz, Victor and Schewenius, Maria and Donges, Jonathan F and Fetzer, Ingo and Zhivkoplias, Erik and Barfuss, Wolfram and Delannoy, Louis and Wang-Erlandsson, Lan and Gelbrecht, Maximilian and Heitzig, Jobst and others},
  journal={arXiv preprint arXiv:2510.24373},
  year={2025}
}

@article{zaidan2025artificial,
  title={Artificial Intelligence for Atmospheric Sciences: A Research Roadmap},
  author={Zaidan, Martha Arbayani and Motlagh, Naser Hossein and Nurmi, Petteri and Hussein, Tareq and Kulmala, Markku and Pet{\"a}j{\"a}, Tuukka and Tarkoma, Sasu},
  journal={arXiv preprint arXiv:2506.16281},
  year={2025}
}

@online{satariano2025aicompute,
  author = {Satariano, Adam and Mozur, Paul},
  title = {A.I. Computing Power Is Splitting the World Into Haves and Have-Nots},
  organization = {The New York Times},
  url = {https://www.nytimes.com/interactive/2025/06/23/technology/ai-computing-global-divide.html},
  day = {23},
  month = {jun},
  year = {2025},
  urldate = {2025-11-19}
}

@article{zha2025data,
  title={Data-centric artificial intelligence: A survey},
  author={Zha, Daochen and Bhat, Zaid Pervaiz and Lai, Kwei-Herng and Yang, Fan and Jiang, Zhimeng and Zhong, Shaochen and Hu, Xia},
  journal={ACM Computing Surveys},
  volume={57},
  number={5},
  pages={1--42},
  year={2025},
  publisher={ACM New York, NY}
}

@article{polasky2025discrepancies,
  title={Discrepancies in precipitation trends between observational and reanalysis datasets in the Amazon Basin},
  author={Polasky, Andrew and Sapkota, Vikrant and Forest, Chris E and Fuentes, Jose D},
  journal={Scientific reports},
  volume={15},
  number={1},
  pages={7268},
  year={2025},
  publisher={Nature Publishing Group UK London}
}

@article{da2024discrepancies,
  title={Discrepancies between observation and ERA5 reanalysis in the Amazon deforestation context: A case study},
  author={da Silva, Queren Priscila and Moreira, Demerval Soares and de Freitas, Helber Cust{\'o}dio and Domingues, Leonardo Moreno},
  journal={Dynamics of Atmospheres and Oceans},
  volume={106},
  pages={101442},
  year={2024},
  publisher={Elsevier}
}

@misc{jiao2021evaluation,
  title={Evaluation of spatial-temporal variation performance of ERA5 precipitation data in China. Sci. Rep., 11, 17956},
  author={Jiao, D and Xu, N and Yang, F and Xu, K},
  year={2021}
}

@article{zou2022performance,
  title={Performance of air temperature from ERA5-Land reanalysis in coastal urban agglomeration of Southeast China},
  author={Zou, Jin and Lu, Ning and Jiang, Hou and Qin, Jun and Yao, Ling and Xin, Ying and Su, Fenzhen},
  journal={Science of The Total Environment},
  volume={828},
  pages={154459},
  year={2022},
  publisher={Elsevier}
}

@article{xin2022evaluation,
  title={Evaluation of IMERG and ERA5 precipitation products over the Mongolian Plateau},
  author={Xin, Ying and Yang, Yaping and Chen, Xiaona and Yue, Xiafang and Liu, Yangxiaoyue and Yin, Cong},
  journal={Scientific reports},
  volume={12},
  number={1},
  pages={21776},
  year={2022},
  publisher={Nature Publishing Group UK London}
}

@article{dueben2022challenges,
  title={Challenges and benchmark datasets for machine learning in the atmospheric sciences: Definition, status, and outlook},
  author={Dueben, Peter D and Schultz, Martin G and Chantry, Matthew and Gagne, David John and Hall, David Matthew and McGovern, Amy},
  journal={Artificial Intelligence for the Earth Systems},
  volume={1},
  number={3},
  pages={e210002},
  year={2022}
}

@article{linsenmeier2023global,
  title={Global inequalities in weather forecasts},
  author={Linsenmeier, Manuel and Shrader, Jeffrey G},
  journal={SocArXiv 7e2jf, Center for Open Science. You might wonder how this can be true when the ECMWF forecasts have so quickly improved in both the Northern and Southern Hemisphere. This is because most countries need to develop their own forecasts to get more high-resolution predictions},
  year={2023}
}

@article{olivetti2025weather,
  author = {Olivetti, Leonardo and Messori, Gabriele},
  title = {Whose weather is it? A fairness framework for data-driven weather forecasting},
  journal = {ESS Open Archive},
  year = {2025},
  month = {oct},
  day = {12},
  doi = {10.22541/essoar.174802937.77365288/v2},
  eprinttype = {essoar},
  eprint = {174802937.77365288/v2}
}

@article{metcalfe2025gaps,
  author = {Metcalfe, Daniel and Anders, Emily and Axén, Hanna and others},
  title = {Gaps in tropical science arising from biased spatial patterns of sampling and citation},
  journal = {Research Square},
  year = {2025},
  month = {jun},
  day = {23},
  doi = {10.21203/rs.3.rs-6829857/v1},
  note = {Preprint (Version 1)}
}

@article{devries2023robust,
  author = {de Vries, I. E. and Sippel, S. and Pendergrass, A. G. and Knutti, R.},
  title = {Robust global detection of forced changes in mean and extreme precipitation despite observational disagreement on the magnitude of change},
  journal = {Earth System Dynamics},
  volume = {14},
  pages = {81--100},
  year = {2023},
  doi = {10.5194/esd-14-81-2023},
  url = {https://doi.org/10.5194/esd-14-81-2023}
}

@article{cattelan2025unraveling,
  title={Unraveling divergences: disagreement between precipitation datasets and stations in tropical South America},
  author={Cattelan, Lu{\'\i}s Gustavo and Mattos, Caio RC and Hirota, Marina},
  journal={Environmental Research Letters},
  volume={20},
  number={10},
  pages={104043},
  year={2025},
  publisher={IOP Publishing}
}

@article{pastorello2020fluxnet2015,
  title={The FLUXNET2015 dataset and the ONEFlux processing pipeline for eddy covariance data},
  author={Pastorello, Gilberto and Trotta, Carlo and Canfora, Eleonora and Chu, Housen and Christianson, Danielle and Cheah, You-Wei and Poindexter, Cristina and Chen, Jiquan and Elbashandy, Abdelrahman and Humphrey, Marty and others},
  journal={Scientific data},
  volume={7},
  number={1},
  pages={225},
  year={2020},
  publisher={Nature Publishing Group UK London}
}

@article{delwiche2021fluxnet,
  author = {Delwiche, K. B. and Knox, S. H. and Malhotra, A. and Fluet-Chouinard, E. and McNicol, G. and Feron, S. and Ouyang, Z. and Papale, D. and Trotta, C. and Canfora, E. and Cheah, Y.-W. and Christianson, D. and Alberto, Ma. C. R. and Alekseychik, P. and Aurela, M. and Baldocchi, D. and Bansal, S. and Billesbach, D. P. and Bohrer, G. and Bracho, R. and Buchmann, N. and Campbell, D. I. and Celis, G. and Chen, J. and Chen, W. and Chu, H. and Dalmagro, H. J. and Dengel, S. and Desai, A. R. and Detto, M. and Dolman, H. and Eichelmann, E. and Euskirchen, E. and Famulari, D. and Fuchs, K. and Goeckede, M. and Gogo, S. and Gondwe, M. J. and Goodrich, J. P. and Gottschalk, P. and Graham, S. L. and Heimann, M. and Helbig, M. and Helfter, C. and Hemes, K. S. and Hirano, T. and Hollinger, D. and Hörtnagl, L. and Iwata, H. and Jacotot, A. and Jurasinski, G. and Kang, M. and Kasak, K. and King, J. and Klatt, J. and Koebsch, F. and Krauss, K. W. and Lai, D. Y. F. and Lohila, A. and Mammarella, I. and Belelli Marchesini, L. and Manca, G. and Matthes, J. H. and Maximov, T. and Merbold, L. and Mitra, B. and Morin, T. H. and Nemitz, E. and Nilsson, M. B. and Niu, S. and Oechel, W. C. and Oikawa, P. Y. and Ono, K. and Peichl, M. and Peltola, O. and Reba, M. L. and Richardson, A. D. and Riley, W. and Runkle, B. R. K. and Ryu, Y. and Sachs, T. and Sakabe, A. and Sanchez, C. R. and Schuur, E. A. and Schäfer, K. V. R. and Sonnentag, O. and Sparks, J. P. and Stuart-Haëntjens, E. and Sturtevant, C. and Sullivan, R. C. and Szutu, D. J. and Thom, J. E. and Torn, M. S. and Tuittila, E.-S. and Turner, J. and Ueyama, M. and Valach, A. C. and Vargas, R. and Varlagin, A. and Vazquez-Lule, A. and Verfaillie, J. G. and Vesala, T. and Vourlitis, G. L. and Ward, E. J. and Wille, C. and Wohlfahrt, G. and Wong, G. X. and Zhang, Z. and Zona, D. and Windham-Myers, L. and Poulter, B. and Jackson, R. B.},
  title = {{FLUXNET-CH4}: a global, multi-ecosystem dataset and analysis of methane seasonality from freshwater wetlands},
  journal = {Earth System Science Data},
  volume = {13},
  pages = {3607--3689},
  year = {2021},
  doi = {10.5194/essd-13-3607-2021},
  url = {https://doi.org/10.5194/essd-13-3607-2021}
}

@article{jung2019fluxcom,
  title={The FLUXCOM ensemble of global land-atmosphere energy fluxes},
  author={Jung, Martin and Koirala, Sujan and Weber, Ulrich and Ichii, Kazuhito and Gans, Fabian and Camps-Valls, Gustau and Papale, Dario and Schwalm, Christopher and Tramontana, Gianluca and Reichstein, Markus},
  journal={Scientific data},
  volume={6},
  number={1},
  pages={74},
  year={2019},
  publisher={Nature Publishing Group UK London}
}

@article{nathaniel2023metaflux,
  title={MetaFlux: Meta-learning global carbon fluxes from sparse spatiotemporal observations},
  author={Nathaniel, Juan and Liu, Jiangong and Gentine, Pierre},
  journal={Scientific Data},
  volume={10},
  number={1},
  pages={440},
  year={2023},
  publisher={Nature Publishing Group UK London}
}

@article{yuan2025global,
  title={Global carbon flux dataset generated by fusing remote sensing and multiple flux networks observation},
  author={Yuan, Qiwang and Wang, Xufeng and Che, Tao and Li, Jun},
  journal={Scientific Data},
  volume={12},
  number={1},
  pages={1359},
  year={2025},
  publisher={Nature Publishing Group UK London}
}

@article{kraemer2025artificial,
  title={Artificial intelligence for modelling infectious disease epidemics},
  author={Kraemer, Moritz UG and Tsui, Joseph L-H and Chang, Serina Y and Lytras, Spyros and Khurana, Mark P and Vanderslott, Samantha and Bajaj, Sumali and Scheidwasser, Neil and Curran-Sebastian, Jacob Liam and Semenova, Elizaveta and others},
  journal={Nature},
  volume={638},
  number={8051},
  pages={623--635},
  year={2025},
  publisher={Nature Publishing Group UK London}
}

@article{wong2023leveraging,
  title={Leveraging artificial intelligence in the fight against infectious diseases},
  author={Wong, Felix and de la Fuente-Nunez, Cesar and Collins, James J},
  journal={Science},
  volume={381},
  number={6654},
  pages={164--170},
  year={2023},
  publisher={American Association for the Advancement of Science}
}

@article{brownstein2023advances,
  title={Advances in artificial intelligence for infectious-disease surveillance},
  author={Brownstein, John S and Rader, Benjamin and Astley, Christina M and Tian, Huaiyu},
  journal={New England Journal of Medicine},
  volume={388},
  number={17},
  pages={1597--1607},
  year={2023},
  publisher={Mass Medical Soc}
}

@article{keshavamurthy2022predicting,
  title={Predicting infectious disease for biopreparedness and response: A systematic review of machine learning and deep learning approaches},
  author={Keshavamurthy, Ravikiran and Dixon, Samuel and Pazdernik, Karl T and Charles, Lauren E},
  journal={One Health},
  volume={15},
  pages={100439},
  year={2022},
  publisher={Elsevier}
}

@article{bremer2017co,
  title={Co-production in climate change research: reviewing different perspectives},
  author={Bremer, Scott and Meisch, Simon},
  journal={Wiley Interdisciplinary Reviews: Climate Change},
  volume={8},
  number={6},
  pages={e482},
  year={2017},
  publisher={Wiley Online Library}
}

@article{turnhout2020politics,
  title={The politics of co-production: participation, power, and transformation},
  author={Turnhout, Esther and Metze, Tamara and Wyborn, Carina and Klenk, Nicole and Louder, Elena},
  journal={Current opinion in environmental sustainability},
  volume={42},
  pages={15--21},
  year={2020},
  publisher={Elsevier}
}

@article{hanigan2006comparison,
  title={A comparison of methods for calculating population exposure estimates of daily weather for health research},
  author={Hanigan, Ivan and Hall, Gillian and Dear, Keith BG},
  journal={International Journal of Health Geographics},
  volume={5},
  number={1},
  pages={38},
  year={2006},
  publisher={Springer}
}

@article{masi2025safe,
  title={SAFE: A Novel Approach to AI Weather Evaluation through Stratified Assessments of Forecasts over Earth},
  author={Masi, Nick and Balestriero, Randall},
  journal={arXiv preprint arXiv:2510.26099},
  year={2025}
}

@article{bodnar2025foundation,
  author = {Bodnar, C. and Bruinsma, W. P. and Lucic, A. and others},
  title = {A foundation model for the Earth system},
  journal = {Nature},
  volume = {641},
  pages = {1180--1187},
  year = {2025},
  doi = {10.1038/s41586-025-09005-y},
  url = {https://doi.org/10.1038/s41586-025-09005-y}
}

@article{allen2025endtoend,
  author = {Allen, A. and Markou, S. and Tebbutt, W. and others},
  title = {End-to-end data-driven weather prediction},
  journal = {Nature},
  volume = {641},
  pages = {1172--1179},
  year = {2025},
  doi = {10.1038/s41586-025-08897-0},
  url = {https://doi.org/10.1038/s41586-025-08897-0}
}

@article{price2025probabilistic,
  author = {Price, I. and Sanchez-Gonzalez, A. and Alet, F. and others},
  title = {Probabilistic weather forecasting with machine learning},
  journal = {Nature},
  volume = {637},
  pages = {84--90},
  year = {2025},
  doi = {10.1038/s41586-024-08252-9},
  url = {https://doi.org/10.1038/s41586-024-08252-9}
}

@article{chen2025operational,
  author = {Chen, K. and Han, T. and Ling, F. and others},
  title = {The operational medium-range deterministic weather forecasting can be extended beyond a 10-day lead time},
  journal = {Communications Earth \& Environment},
  volume = {6},
  pages = {518},
  year = {2025},
  doi = {10.1038/s43247-025-02502-y},
  url = {https://doi.org/10.1038/s43247-025-02502-y}
}

@article{lam2023learning,
  author = {Lam, R\'emi and others},
  title = {Learning skillful medium-range global weather forecasting},
  journal = {Science},
  volume = {382},
  pages = {1416--1421},
  year = {2023},
  doi = {10.1126/science.adi2336},
  url = {https://doi.org/10.1126/science.adi2336}
}

@article{lang2412aifs,
  title={AIFS-CRPS: Ensemble forecasting using a model trained with a loss function based on the continuous ranked probability score, 2024b},
  author={Lang, Simon and Alexe, Mihai and Clare, Mariana CA and Roberts, Christopher and Adewoyin, Rilwan and Bouall{\`e}gue, Zied Ben and Chantry, Matthew and Dramsch, Jesper and Dueben, Peter D and Hahner, Sara and others},
  journal={URL https://arxiv. org/abs/2412.15832}
}

@misc{lang2024aifsecmwfsdatadriven,
      title={AIFS -- ECMWF's data-driven forecasting system}, 
      author={Simon Lang and Mihai Alexe and Matthew Chantry and Jesper Dramsch and Florian Pinault and Baudouin Raoult and Mariana C. A. Clare and Christian Lessig and Michael Maier-Gerber and Linus Magnusson and Zied Ben Bouallègue and Ana Prieto Nemesio and Peter D. Dueben and Andrew Brown and Florian Pappenberger and Florence Rabier},
      year={2024},
      eprint={2406.01465},
      archivePrefix={arXiv},
      primaryClass={physics.ao-ph},
      url={https://arxiv.org/abs/2406.01465}, 
}

@article{bi2023accurate,
  author = {Bi, K. and Xie, L. and Zhang, H. and others},
  title = {Accurate medium-range global weather forecasting with 3D neural networks},
  journal = {Nature},
  volume = {619},
  pages = {533--538},
  year = {2023},
  doi = {10.1038/s41586-023-06185-3},
  url = {https://doi.org/10.1038/s41586-023-06185-3}
}

@misc{pathak2022fourcastnetglobaldatadrivenhighresolution,
      title={FourCastNet: A Global Data-driven High-resolution Weather Model using Adaptive Fourier Neural Operators}, 
      author={Jaideep Pathak and Shashank Subramanian and Peter Harrington and Sanjeev Raja and Ashesh Chattopadhyay and Morteza Mardani and Thorsten Kurth and David Hall and Zongyi Li and Kamyar Azizzadenesheli and Pedram Hassanzadeh and Karthik Kashinath and Animashree Anandkumar},
      year={2022},
      eprint={2202.11214},
      archivePrefix={arXiv},
      primaryClass={physics.ao-ph},
      url={https://arxiv.org/abs/2202.11214}, 
}

@misc{nguyen2023climaxfoundationmodelweather,
      title={ClimaX: A foundation model for weather and climate}, 
      author={Tung Nguyen and Johannes Brandstetter and Ashish Kapoor and Jayesh K. Gupta and Aditya Grover},
      year={2023},
      eprint={2301.10343},
      archivePrefix={arXiv},
      primaryClass={cs.LG},
      url={https://arxiv.org/abs/2301.10343}, 
}

@article{sun2025data,
  author = {Sun, X. and Zhong, X. and Xu, X. and others},
  title = {A data-to-forecast machine learning system for global weather},
  journal = {Nature Communications},
  volume = {16},
  pages = {6658},
  year = {2025},
  doi = {10.1038/s41467-025-62024-1},
  url = {https://doi.org/10.1038/s41467-025-62024-1}
}

@misc{kaltenborn2023climatesetlargescaleclimatemodel,
      title={ClimateSet: A Large-Scale Climate Model Dataset for Machine Learning}, 
      author={Julia Kaltenborn and Charlotte E. E. Lange and Venkatesh Ramesh and Philippe Brouillard and Yaniv Gurwicz and Chandni Nagda and Jakob Runge and Peer Nowack and David Rolnick},
      year={2023},
      eprint={2311.03721},
      archivePrefix={arXiv},
      primaryClass={cs.LG},
      url={https://arxiv.org/abs/2311.03721}, 
}

@article{Rasp_2020,
   title={WeatherBench: A Benchmark Data Set for Data‐Driven Weather Forecasting},
   volume={12},
   ISSN={1942-2466},
   url={http://dx.doi.org/10.1029/2020MS002203},
   DOI={10.1029/2020ms002203},
   number={11},
   journal={Journal of Advances in Modeling Earth Systems},
   publisher={American Geophysical Union (AGU)},
   author={Rasp, Stephan and Dueben, Peter D. and Scher, Sebastian and Weyn, Jonathan A. and Mouatadid, Soukayna and Thuerey, Nils},
   year={2020},
   month=nov }

@misc{rasp2024weatherbench2benchmarkgeneration,
      title={WeatherBench 2: A benchmark for the next generation of data-driven global weather models}, 
      author={Stephan Rasp and Stephan Hoyer and Alexander Merose and Ian Langmore and Peter Battaglia and Tyler Russel and Alvaro Sanchez-Gonzalez and Vivian Yang and Rob Carver and Shreya Agrawal and Matthew Chantry and Zied Ben Bouallegue and Peter Dueben and Carla Bromberg and Jared Sisk and Luke Barrington and Aaron Bell and Fei Sha},
      year={2024},
      eprint={2308.15560},
      archivePrefix={arXiv},
      primaryClass={physics.ao-ph},
      url={https://arxiv.org/abs/2308.15560}, 
}

@misc{ran2025hrextremehighresolutiondatasetextreme,
      title={HR-Extreme: A High-Resolution Dataset for Extreme Weather Forecasting}, 
      author={Nian Ran and Peng Xiao and Yue Wang and Wesley Shi and Jianxin Lin and Qi Meng and Richard Allmendinger},
      year={2025},
      eprint={2409.18885},
      archivePrefix={arXiv},
      primaryClass={cs.LG},
      url={https://arxiv.org/abs/2409.18885}, 
}

@misc{yu2025climdetectbenchmarkdatasetclimate,
      title={ClimDetect: A Benchmark Dataset for Climate Change Detection and Attribution}, 
      author={Sungduk Yu and Brian L. White and Anahita Bhiwandiwalla and Musashi Hinck and Matthew Lyle Olson and Yaniv Gurwicz and Raanan Y. Rohekar and Tung Nguyen and Vasudev Lal},
      year={2025},
      eprint={2408.15993},
      archivePrefix={arXiv},
      primaryClass={cs.CV},
      url={https://arxiv.org/abs/2408.15993}, 
}

@article{watson2022climatebench,
  title={ClimateBench v1. 0: A benchmark for data-driven climate projections},
  author={Watson-Parris, Duncan and Rao, Yuhan and Olivi{\'e}, Dirk and Seland, {\O}yvind and Nowack, Peer and Camps-Valls, Gustau and Stier, Philip and Bouabid, Shahine and Dewey, Maura and Fons, Emilie and others},
  journal={Journal of Advances in Modeling Earth Systems},
  volume={14},
  number={10},
  pages={e2021MS002954},
  year={2022},
  publisher={Wiley Online Library}
}

@misc{fu2025climatebenchmmultimodalclimatedata,
      title={ClimateBench-M: A Multi-Modal Climate Data Benchmark with a Simple Generative Method}, 
      author={Dongqi Fu and Yada Zhu and Zhining Liu and Lecheng Zheng and Xiao Lin and Zihao Li and Liri Fang and Katherine Tieu and Onkar Bhardwaj and Kommy Weldemariam and Hanghang Tong and Hendrik Hamann and Jingrui He},
      year={2025},
      eprint={2504.07394},
      archivePrefix={arXiv},
      primaryClass={cs.LG},
      url={https://arxiv.org/abs/2504.07394}, 
}

@misc{shinde2024wxcbenchnoveldatasetweather,
      title={WxC-Bench: A Novel Dataset for Weather and Climate Downstream Tasks}, 
      author={Rajat Shinde and Christopher E. Phillips and Kumar Ankur and Aman Gupta and Simon Pfreundschuh and Sujit Roy and Sheyenne Kirkland and Vishal Gaur and Amy Lin and Aditi Sheshadri and Udaysankar Nair and Manil Maskey and Rahul Ramachandran},
      year={2024},
      eprint={2412.02780},
      archivePrefix={arXiv},
      primaryClass={cs.LG},
      url={https://arxiv.org/abs/2412.02780}, 
}

@misc{yu2024climsimonlinelargemultiscaledataset,
      title={ClimSim-Online: A Large Multi-scale Dataset and Framework for Hybrid ML-physics Climate Emulation}, 
      author={Sungduk Yu and Zeyuan Hu and Akshay Subramaniam and Walter Hannah and Liran Peng and Jerry Lin and Mohamed Aziz Bhouri and Ritwik Gupta and Björn Lütjens and Justus C. Will and Gunnar Behrens and Julius J. M. Busecke and Nora Loose and Charles I. Stern and Tom Beucler and Bryce Harrop and Helge Heuer and Benjamin R. Hillman and Andrea Jenney and Nana Liu and Alistair White and Tian Zheng and Zhiming Kuang and Fiaz Ahmed and Elizabeth Barnes and Noah D. Brenowitz and Christopher Bretherton and Veronika Eyring and Savannah Ferretti and Nicholas Lutsko and Pierre Gentine and Stephan Mandt and J. David Neelin and Rose Yu and Laure Zanna and Nathan Urban and Janni Yuval and Ryan Abernathey and Pierre Baldi and Wayne Chuang and Yu Huang and Fernando Iglesias-Suarez and Sanket Jantre and Po-Lun Ma and Sara Shamekh and Guang Zhang and Michael Pritchard},
      year={2024},
      eprint={2306.08754},
      archivePrefix={arXiv},
      primaryClass={cs.LG},
      url={https://arxiv.org/abs/2306.08754}, 
}

@article{demaeyer2023euppbench,
  title={The EUPPBench postprocessing benchmark dataset v1. 0},
  author={Demaeyer, Jonathan and Bhend, Jonas and Lerch, Sebastian and Primo, Cristina and Van Schaeybroeck, Bert and Atencia, Aitor and Ben Bouall{\`e}gue, Zied and Chen, Jieyu and Dabernig, Markus and Evans, Gavin and others},
  journal={Earth System Science Data},
  volume={15},
  number={6},
  pages={2635--2653},
  year={2023},
  publisher={Copernicus GmbH}
}

@misc{racah2017extremeweatherlargescaleclimatedataset,
      title={ExtremeWeather: A large-scale climate dataset for semi-supervised detection, localization, and understanding of extreme weather events}, 
      author={Evan Racah and Christopher Beckham and Tegan Maharaj and Samira Ebrahimi Kahou and Prabhat and Christopher Pal},
      year={2017},
      eprint={1612.02095},
      archivePrefix={arXiv},
      primaryClass={cs.CV},
      url={https://arxiv.org/abs/1612.02095}, 
}

@article{prabhat2020climatenet,
  title={ClimateNet: an expert-labelled open dataset and deep learning architecture for enabling high-precision analyses of extreme weather},
  author={Prabhat and Kashinath, Karthik and Mudigonda, Mayur and Kim, Sol and Kapp-Schwoerer, Lukas and Graubner, Andre and Karaismailoglu, Ege and von Kleist, Leo and Kurth, Thorsten and Greiner, Annette and others},
  journal={Geoscientific Model Development Discussions},
  volume={2020},
  pages={1--28},
  year={2020},
  publisher={G{\"o}ttingen, Germany}
}

@misc{zhao2025exebenchbenchmarkingfoundationmodels,
      title={ExEBench: Benchmarking Foundation Models on Extreme Earth Events}, 
      author={Shan Zhao and Zhitong Xiong and Jie Zhao and Xiao Xiang Zhu},
      year={2025},
      eprint={2505.08529},
      archivePrefix={arXiv},
      primaryClass={cs.LG},
      url={https://arxiv.org/abs/2505.08529}, 
}

@article { IdentifyingandCategorizingBiasinAIMLforEarthSciences,
      author = "Amy McGovern and Ann Bostrom and Marie McGraw and Randy J. Chase and David John Gagne and Imme Ebert-Uphoff and Kate D. Musgrave and Andrea Schumacher",
      title = "Identifying and Categorizing Bias in AI/ML for Earth Sciences",
      journal = "Bulletin of the American Meteorological Society",
      year = "2024",
      publisher = "American Meteorological Society",
      address = "Boston MA, USA",
      volume = "105",
      number = "3",
      doi = "10.1175/BAMS-D-23-0196.1",
      pages=      "E567 - E583",
      url = "https://journals.ametsoc.org/view/journals/bams/105/3/BAMS-D-23-0196.1.xml"
}

@misc{subich2025fixingdoublepenaltydatadriven,
      title={Fixing the Double Penalty in Data-Driven Weather Forecasting Through a Modified Spherical Harmonic Loss Function}, 
      author={Christopher Subich and Syed Zahid Husain and Leo Separovic and Jing Yang},
      year={2025},
      eprint={2501.19374},
      archivePrefix={arXiv},
      primaryClass={cs.LG},
      url={https://arxiv.org/abs/2501.19374}, 
}

@article{
doi:10.1126/science.aec8146,
author = {Petteri Taalas },
title = {Free global access to climate and weather data must continue},
journal = {Science},
volume = {390},
number = {6769},
pages = {111-111},
year = {2025},
doi = {10.1126/science.aec8146},
URL = {https://www.science.org/doi/abs/10.1126/science.aec8146},
eprint = {https://www.science.org/doi/pdf/10.1126/science.aec8146}}

@article{MADDAH2023,
title = {Effectiveness of Public Health Digital Surveillance Systems for Infectious Disease Prevention and Control at Mass Gatherings: Systematic Review},
journal = {Journal of Medical Internet Research},
volume = {25},
year = {2023},
issn = {1438-8871},
doi = {https://doi.org/10.2196/44649},
url = {https://www.sciencedirect.com/science/article/pii/S1438887123003722},
author = {Noha Maddah and Arpana Verma and Maryam Almashmoum and John Ainsworth}}

@article{Schulze2023,
title = {Digital dashboards visualizing public health data: a systematic review},
journal = {Frontiers in Public Health.},
volume = {11},
year = {2023},
issn = {999958},
doi = {https://doi.org/10.3389/fpubh.2023.999958},
thor = {Schulze, A. and Brand, F. and Geppert, J. and Bol, G. F.}}

@misc{roser2024,
author = {Max Roser and Hannah Ritchie and Fiona Spooner},
title = {Burden of Disease},
  organization = {Our World in Data},
  url = {https://ourworldindata.org/burden-of-disease},
  day = {24},
  month = {nov},
  year = {2025},
  urldate = {2024-02-01}
}

@article{regilme2024artificial,
  title={Artificial intelligence colonialism: Environmental damage, labor exploitation, and human rights crises in the Global South},
  author={Regilme, Salvador Santino F},
  journal={SAIS Review of International Affairs},
  volume={44},
  number={2},
  pages={75--92},
  year={2024},
  publisher={Johns Hopkins University Press}
}

@techreport{lemos2025digital,
  title       = {Digital Public Infrastructure for Climate: The Missing Backbone for Climate Action},
  author      = {Lemos, Ronaldo},
  institution = {Institute for Technology and Society of Rio de Janeiro (ITS Rio)},
  year        = {2025},
  month       = {October},
  note        = {Report commissioned by the COP30 Presidency},
  url         = {https://itsrio.org}
}

@article{nature2025energy,
  title={AI’s energy demand is booming — but where and by how much?},
  author={Nature},
  journal={Nature},
  volume={639},
  pages={19--21},
  year={2025},
  publisher={Nature Publishing Group},
  doi={10.1038/d41586-025-00616-z}
}

@article{li2023making,
  title={Making AI less ``thirsty'': The hidden water footprint of AI models},
  author={Li, Pengfei and Yang, Jianyi and Islam, Mohammad A and Ren, Shaolei},
  journal={arXiv preprint arXiv:2304.03271},
  year={2023}
}

@misc{khadir2025democracyainumericalweather,
      title={Democracy of AI Numerical Weather Models: An Example of Global Forecasting with FourCastNetv2 Made by a University Research Lab Using GPU}, 
      author={Iman Khadir and Shane Stevenson and Henry Li and Kyle Krick and Abram Burrows and David Hall and Stan Posey and Samuel S. P. Shen},
      year={2025},
      eprint={2504.17028},
      archivePrefix={arXiv},
      primaryClass={cs.LG},
      url={https://arxiv.org/abs/2504.17028}, 
}

@misc{azhar2025fully,
  author       = {Azhar, Ali},
  title        = {Fully {AI}-Driven System Signals a New Era in Weather Forecasting},
  year         = {2025},
  month        = {mar},
  howpublished = {HPCwire},
  url          = {https://www.hpcwire.com/aiwire/2025/03/25/fully-ai-driven-system-signals-a-new-era-in-weather-forecasting/},
  note         = {Accessed: 2025-12-10}
}

@article{hall2025ai,
  author       = {Hall, Rachel and Sample, Ian},
  title        = {{AI}-driven weather prediction breakthrough reported},
  year         = {2025},
  month        = {mar},
  day          = {20},
  journal      = {The Guardian},
  url          = {https://www.theguardian.com/technology/2025/mar/20/ai-aardvark-weather-prediction-forecasting-artificial-intelligence},
  note         = {Accessed: 2025-12-10}
}

@article{kuznetsov2025transforming,
  title={Transforming climate services with LLMs and multi-source data integration},
  author={Kuznetsov, Ivan and Jost, Antonia Anna and Pantiukhin, Dmitrii and Shapkin, Boris and Jung, Thomas and Koldunov, Nikolay},
  journal={npj Climate Action},
  volume={4},
  number={1},
  pages={97},
  year={2025},
  publisher={Nature Publishing Group UK London}
}

@inproceedings{marion2023when,
    title={When Less is More: Investigating Data Pruning for Pretraining {LLM}s at Scale},
    author={Max Marion and Ahmet Üstün and Luiza Pozzobon and Alex Wang and Marzieh Fadaee and Sara Hooker},
    booktitle={NeurIPS Workshop on Attributing Model Behavior at Scale},
    year={2023},
    url={https://openreview.net/forum?id=XUIYn3jo5T}
}

@inproceedings{zewang2025bias,
    title={Bias Amplification: Large Language Models as Increasingly Biased Media},
    author={Ze Wang and Zekun Wu and Jeremy Zhang and Xin Guan and Navya Jain and Skylar Lu and Saloni Gupta and Adriano Koshiyama},
    booktitle={Submitted to ACL Rolling Review - February 2025},
    year={2025},
    url={https://openreview.net/forum?id=X5IaJtOtNL},
    note={under review}
}

@misc{lin2024designing,
      title={Designing Domain-Specific Large Language Models: The Critical Role of Fine-Tuning in Public Opinion Simulation}, 
      author={Haocheng Lin},
      year={2024},
      eprint={2409.19308},
      archivePrefix={arXiv},
      primaryClass={cs.CL},
      url={https://arxiv.org/abs/2409.19308}, 
}

@article{pasgaard2015geographical,
    title = {Geographical imbalances and divides in the scientific production of climate change knowledge},
    author = {Maya Pasgaard and Bo Dalsgaard and Pietro K. Maruyama and Brody Sandel and Niels Strange},
    journal = {Global Environmental Change},
    volume = {35},
    pages = {279-288},
    year = {2015},
    doi = {https://doi.org/10.1016/j.gloenvcha.2015.09.018}
    }

@article{manvi2024large,
  title={Large language models are geographically biased},
  author={Manvi, Rohin and Khanna, Samar and Burke, Marshall and Lobell, David and Ermon, Stefano},
  journal={arXiv preprint arXiv:2402.02680},
  year={2024}
}

@article{overland2022funding,
  title={Funding flows for climate change research on Africa: where do they come from and where do they go?},
  author={Overland, Indra and Fossum Sagbakken, Haakon and Isataeva, Aidai and Kolodzinskaia, Galina and Simpson, Nicholas Philip and Trisos, Christopher and Vakulchuk, Roman},
  journal={Climate and Development},
  volume={14},
  number={8},
  pages={705--724},
  year={2022},
  publisher={Taylor \& Francis}
}

@inproceedings{sanu2024limitations,
    title={Limitations of Large Language Models},   
    author={Sanu, Erin and Amudaa, T Keerthi and Bhat, Prasiddha and Dinesh, Guduru and Kumar Chate, Apoorva Uday and P, Ramakanth Kumar},
    booktitle={2024 8th International Conference on Computational System and Information Technology for Sustainable Solutions (CSITSS)}, 
    year={2024},
    volume={},
    number={},
    pages={1-6},
    doi={10.1109/CSITSS64042.2024.10817070}
}

@inproceedings{bender2021dangers,
    title = {On the Dangers of Stochastic Parrots: Can Language Models Be Too Big?},
    author = {Bender, Emily M. and Gebru, Timnit and McMillan-Major, Angelina and Shmitchell, Shmargaret},
    year = {2021},
    publisher = {Association for Computing Machinery},
    url = {https://doi.org/10.1145/3442188.3445922},
    doi = {10.1145/3442188.3445922},
    booktitle = {Proceedings of the 2021 ACM Conference on Fairness, Accountability, and Transparency},
    pages = {610–623},
    numpages = {14},
    series = {FAccT '21}
}

@misc{unpress2023gaef3587,
  title        = {United Nations General Assembly Press Release: GA/EF/3587},
  author       = {{United Nations}},
  year         = {2023},
  month        = {November},
  howpublished = {Press release, United Nations},
  url          = {https://press.un.org/en/2023/gaef3587.doc.htm}
}

@article{sambasivan2021everyone,
  title={“Everyone wants to do the model work, not the data work”: Data Cascades in High-Stakes AI},
  author={Nithya Sambasivan and Shivani Kapania and Hannah Highfill and Diana Akrong and Praveen K. Paritosh and Lora Aroyo},
  journal={Proceedings of the 2021 CHI Conference on Human Factors in Computing Systems},
  year={2021},
  url={https://api.semanticscholar.org/CorpusID:231829607}
}

@article{hersbach2020era5,
  title={The ERA5 global reanalysis},
  author={Hersbach, Hans and Bell, Bill and Berrisford, Paul and Hirahara, Shoji and Hor{\'a}nyi, Andr{\'a}s and Mu{\~n}oz-Sabater, Joaqu{\'\i}n and Nicolas, Julien and Peubey, Carole and Radu, Raluca and Schepers, Dinand and others},
  journal={Quarterly journal of the royal meteorological society},
  volume={146},
  number={730},
  pages={1999--2049},
  year={2020},
  publisher={Wiley Online Library}
}

@article{eyring2016overview,
  title={Overview of the Coupled Model Intercomparison Project Phase 6 (CMIP6) experimental design and organization},
  author={Eyring, Veronika and Bony, Sandrine and Meehl, Gerald A and Senior, Catherine A and Stevens, Bjorn and Stouffer, Ronald J and Taylor, Karl E},
  journal={Geoscientific Model Development},
  volume={9},
  number={5},
  pages={1937--1958},
  year={2016},
  publisher={Copernicus GmbH}
}

@article{lavers2022evaluation,
  title={An evaluation of ERA5 precipitation for climate monitoring},
  author={Lavers, David A and Simmons, Adrian and Vamborg, Freja and Rodwell, Mark J},
  journal={Quarterly Journal of the Royal Meteorological Society},
  volume={148},
  number={748},
  pages={3152--3165},
  year={2022},
  publisher={Wiley Online Library}
}

@article{hoffmann2022training,
  title={Training Compute-Optimal Large Language Models},
  author={Hoffmann, Jordan and Borgeaud, Sebastian and Mensch, Arthur and others},
  journal={arXiv preprint arXiv:2203.15556},
  year={2022}
}

@article{eyre2022value,
  title={The value of surface-based meteorological observation data},
  author={Eyre, JR and Kull, DW and Riishojgaard, LP and others},
  journal={World Bank, Washington, DC},
  year={2022},
  note={Available at: \url{https://openknowledge.worldbank.org/handle/10986/35178}}
}

@article{plesiat2024artificial,
  title={Artificial intelligence reveals past climate extremes by reconstructing historical records},
  author={Pl{\'e}siat, {\'E}tienne and Dunn, Robert JH and Donat, Markus G and Kadow, Christopher},
  journal={Nature Communications},
  volume={15},
  number={1},
  pages={9191},
  year={2024},
  publisher={Nature Publishing Group UK London}
}

@article{guo2025reconstructed,
  title={Reconstructed global monthly burned area maps from 1901 to 2020},
  author={Guo, Zhixuan and Li, Wei and Ciais, Philippe and Sitch, Stephen and van der Werf, Guido R and Bowring, Simon PK and Bastos, Ana and Mouillot, Florent and He, Jiaying and Sun, Minxuan and others},
  journal={Earth System Science Data Discussions},
  volume={2025},
  pages={1--28},
  year={2025},
  publisher={G{\"o}ttingen, Germany}
}

\begin{appendices}

\section{Supplementary Tables}\label{secA1}

An appendix contains supplementary information that is not an essential part of the text itself but which may be helpful in providing a more comprehensive understanding of the research problem or it is information that is too cumbersome to be included in the body of the paper.

\begin{table}[htbp]
    \centering
    \scriptsize 
    \setlength{\tabcolsep}{2pt} 
    \renewcommand{\arraystretch}{1.3} 
    \caption{List of the frontier weather and climate models, developers, training data, metrics, and compute cost (Table A1)}
    \label{tab:frontier_models_appendix}
    
    
    \begin{tabularx}{\textwidth}{| >{\raggedright\arraybackslash}p{1.7cm} | c | >{\raggedright\arraybackslash\hsize=0.6\hsize}X | >{\raggedright\arraybackslash\hsize=1.5\hsize}X | >{\raggedright\arraybackslash}p{1.2cm} | >{\raggedright\arraybackslash\hsize=0.9\hsize}X |}
        \hline
        \textbf{Model} & \textbf{Year} & \textbf{Developer} & \textbf{Data used for training} & \textbf{Metrics} & \textbf{Training Compute} \\
        \hline
        
        Aurora \citep{bodnar2025foundation} & 2025 & Microsoft Research & Diverse mix of 6 datasets (ERA5, CMCC, IFS-HR, HRES, GFS, GFS Forecasts) & RMSE, MAE & 2.5 Weeks on 32 A100 \\
        \hline
        
        Ardavak Weather \citep{allen2025endtoend} & 2025 & Univ. of Cambridge, Turing Inst., Microsoft, ECMWF & Raw observations (sat radiances, station data) \& ERA5 data. & RMSE, MAE, ACC & 100 hours on A100 \\
        \hline
        
        WeatherNext GenCast \citep{price2025probabilistic} & 2025 & Google DeepMind + Google Research & ERA5 reanalysis data. & CRPS, Spread & 5 days on 32 TPUv5\\
        \hline
        
        FengWu \citep{chen2025operational} & 2025 & Shanghai AI Lab & ERA5 reanalysis data. & RMSE, ACC & not reported  \\
        \hline
        
        WeatherNext GraphCast \citep{lam2023learning} & 2024 & Google DeepMind + Google Research & 39 years (1979–2017) of ERA5 reanalysis data. & RMSE \& ACC & 4 weeks  on 32 TPU \\
        \hline
        
        NeuralGCM \citep{kochkov2024neural} & 2024 & Google, MIT, Harvard \& ECMWF & ERA5 reanalysis data. & RMSE, RMSB,  CRPS, spread-skill ratio &  Not reported  \\
        \hline
        
        AIFS-CRPS \citep{lang2412aifs} & 2024 & ECMWF & Pre-trained on ERA5; fine-tuned on ECMWF operational. & CRPS, RMSE  & low -res. for 4 days on 64 H100 and high-res. for 7 days on 128 H100 \\
        \hline
        
        AIFS \citep{lang2024aifsecmwfsdatadriven} & 2024 & ECMWF & Pre-trained on ERA5; fine-tuned on ECMWF (HRES). & RMSE, ACC & 1 week on 64 A100 \\
        \hline
        
        Pangu-weather \citep{bi2023accurate} & 2023 & Huawei (Cloud) & 43 years (1979-2021) of hourly ERA5 reanalysis data. & RMSE \& ACC & 16 days per model on 192 V100 \\
        \hline
        
        FourCastNet (v1, v2, v3) \citep{pathak2022fourcastnetglobaldatadrivenhighresolution,bonev2025fourcastnet} & 2022, 2025 & NVIDIA \& collaborators & 40 years of ERA5 reanalysis data at 0.25° resolution. & RMSE \& ACC & V1: 16 hours   onc64 A100  V3: three-stage training with 78 hours on 1024, 15 hours on 512 and 8 hours on 256 H100 \\
        \hline

        ClimaX \citep{nguyen2023climaxfoundationmodelweather} & 2023 & Microsoft Research \& UCLA & Pre-trained on CMIP6, fine-tuned on ERA5. & RMSE \& ACC & pretrained on 80 V100 (training time not mentioned), and fine-tuned on 8 V100 for 15 hours per set of weights \\
        \hline
        
        FuXI Weather \citep{sun2025data} & 2023 & Fudan Univ., CMA & 39 years (1979–2017) of ERA5 reanalysis data. & RMSE, ACC & not reported \\
        \hline
    \end{tabularx}
\end{table}

\begin{table}[htbp]
    \centering
    \small 
    \renewcommand{\arraystretch}{1.5} 
    \caption{List of the most common weather and climate benchmark datasets (Table A2)}
    \label{tab:benchmark_datasets_appendix}
    
    \begin{tabularx}{\textwidth}{|l|c| >{\RaggedRight}X |}
        \hline
        \textbf{Benchmark name} & \textbf{Year} & \textbf{Datasets that are included} \\
        \hline
        
        ClimateSet \citep{kaltenborn2023climatesetlargescaleclimatemodel}& 2023 & Input4MIPs, CMIP6 (36 climate models) \\
        \hline
        
        Weatherbench 1 \citep{Rasp_2020} & 2020 & ERA5 \\
        \hline
        
        Weatherbench 2 \citep{rasp2024weatherbench2benchmarkgeneration}& 2023 & ERA5, IFS Operational Forecasts, GFS \\
        \hline
        
        HR-Extreme \citep{ran2025hrextremehighresolutiondatasetextreme}& 2024 & High-Resolution Rapid Refresh (HRRR), NOAA Storm Events \\
        \hline

        ExEBench \citep{zhao2025exebenchbenchmarkingfoundationmodels}& 2025 & CMIP6, ERA5 \\
        \hline
        
        ClimDetect \citep{yu2025climdetectbenchmarkdatasetclimate}& 2023 & CMIP6, ERA5, JRA-3Q, MERRA-2 \\
        \hline
        
        ClimateBench \citep{watson2022climatebench}& 2021 & NorESM2, CMIP6 \\
        \hline
        
        ClimateBench-M \citep{fu2025climatebenchmmultimodalclimatedata}& 2024 & ERA5, NOAA extreme weather data, NASA HLS \\
        \hline
        
        WxC-Bench \citep{shinde2024wxcbenchnoveldatasetweather}& 2024 & ERA5, CMIP6 \\
        \hline
        
        ClimSim-Online \citep{yu2024climsimonlinelargemultiscaledataset}& 2023 & E3SMv1 \\
        \hline
        
        EUPPBench \citep{demaeyer2023euppbench}& 2023 & ECMWF ensemble forecasts, ERA5 \\
        \hline
        
        ExtremeWeather \citep{racah2017extremeweatherlargescaleclimatedataset}& 2017 & ERA5 - labelled for extreme events \\
        \hline
        
        ClimateNet \citep{prabhat2020climatenet}& 2021 & CAM5.1 - labelled for tropical cyclones and atmospheric rivers \\
        \hline
        
    \end{tabularx}
\end{table}

\begin{table}[htbp]
    \centering
    \caption{List of selected HPC Centres and Systems for Weather and Climate Modeling (Table A3) adopted and expanded from \citep{zaidan2025artificial}}
    \label{tab:hpc_countries_appendix}
    \begin{tabular}{llll}
        \toprule
        \textbf{Continent} & \textbf{Institution} & \textbf{Country / Location} & \textbf{System Name(s)} \\
        \midrule
        Europe & BSC-CNS & Spain (Barcelona) & ``MareNostrum 5'' \\
        Europe & CINECA & Italy (Bologna) & ``Leonardo'' \\
        Europe & DKRZ & Germany (Hamburg) & ``Levante'' \\
        Europe & DWD & Germany (Offenbach) & NEC SX-Aurora TSUBASA \\
        Europe & ECMWF & Italy (Bologna) & Atos BullSequana XH2000 \\
        Europe & CSC & Finland (Kajaani) & ``LUMI'' \\
        Europe & Météo-France & France (Toulouse) & Atos/Eviden BullSequana XH2000 \\
        Europe & UKMO & United Kingdom & Microsoft Azure HPC (Cloud) \\
        \midrule
        N. America & DOE (ANL) & USA (Lemont, IL) & ``Chrysalis'' \\
        N. America & DOE (LBNL) & USA (Berkeley, CA) & ``Perlmutter'' \\
        N. America & DOE (ORNL) & USA (Oak Ridge, TN) & ``Frontier'' \\
        N. America & ECCC & Canada (Montréal, QC) & ``Anne Barbara Underhill'' \& ``André Robert'' \\
        N. America & NCAR & USA (Cheyenne, WY) & ``Derecho'' \\
        N. America & NOAA & USA (Various) & WCOSS \& ``Gaea'' \\
        \midrule
        Asia & CMA & China & ``Fengyu'' \\
        Asia & IITM & India (Pune) & ``Arka'' \\
        Asia & JMA & Japan & Fujitsu PRIMEHPC FX1000 \\
        Asia & KMA & South Korea & ``Guru'' \& ``Maru'' \\
        Asia & NCMRWF & India (Noida) & ``Arunika'' \\
        \midrule
        Oceania & NCI & Australia (Canberra) & ``Gadi'' \\
        Oceania & BoM & Australia (Melbourne) & ``Australis II'' \\
        Oceania & NIWA & New Zealand & ``Fitzroy'' \\
        \bottomrule
    \end{tabular}
\end{table}
\end{appendices}

\end{document}